\pdfoutput=1

\documentclass[11pt,twoside,a4paper,cmspaper,final,collab]{cms-tdr}
\def\svnVersion{447294P}\def\cmsCernNoTag{CERN-EP-2017-224}\def\cmsCernDate{\today}\def\cmsMessage{Published in Physics Letters B as \href{http://dx.doi.org/10.1016/j.physletb.2018.01.077}{\doi{10.1016/j.physletb.2018.01.077.}}}

\begin{document}\cmsNoteHeader{B2G-17-003}

\hyphenation{had-ron-i-za-tion}
\hyphenation{cal-or-i-me-ter}
\hyphenation{de-vices}
\RCS$HeadURL: svn+ssh://svn.cern.ch/reps/tdr2/papers/B2G-17-003/trunk/B2G-17-003.tex $
\RCS$Id: B2G-17-003.tex 447271 2018-02-21 17:03:21Z vorobiev $
\newlength\cmsFigWidth
\ifthenelse{\boolean{cms@external}}{\setlength\cmsFigWidth{0.98\columnwidth}}{\setlength\cmsFigWidth{0.75\textwidth}}
\newlength\cmsTabSkip\setlength\cmsTabSkip{1.5ex}
\ifthenelse{\boolean{cms@external}}{\providecommand{\cmsLeft}{top}}{\providecommand{\cmsLeft}{left}}
\ifthenelse{\boolean{cms@external}}{\providecommand{\cmsRight}{bottom}}{\providecommand{\cmsRight}{right}}
\ifthenelse{\boolean{cms@external}}{\setlength\cmsFigWidth{0.95\columnwidth}}{\setlength\cmsFigWidth{0.6\textwidth}}
\ifthenelse{\boolean{cms@external}}{\providecommand{\cmsLeft}{top}}{\providecommand{\cmsLeft}{left}}
\ifthenelse{\boolean{cms@external}}{\providecommand{\cmsRight}{bottom}}{\providecommand{\cmsRight}{right}}
\providecommand{\CL[1]}{\ensuremath{\text{CL}_\text{#1}}\xspace}
\providecommand{\ttjets}{\ensuremath{\ttbar+\text{jets}}\xspace}

\providecommand{\T}{\ensuremath{\mathrm{T}}\xspace}
\providecommand{\Y}{\ensuremath{\mathrm{Y}}\xspace}
\providecommand{\Tbar}{\ensuremath{\overline{\mathrm{T}}}}
\providecommand{\TTbar}{\ensuremath{\mathrm{T}\overline{\mathrm{T}}}\xspace}
\providecommand{\YYbar}{\ensuremath{\mathrm{Y}\overline{\mathrm{Y}}}\xspace}
\providecommand{\bW}{\ensuremath{\PQb\PW}\xspace}
\providecommand{\bWbW}{\ensuremath{\PQb\PW\PAQb\PW}\xspace}
\providecommand{\Mreco}{\ensuremath{m_{\text{reco}}}}
\providecommand{\Trm}{\ensuremath{\mathrm{T}}}
\providecommand{\Lrm}{\ensuremath{\mathrm{L}}}
\providecommand{\pL}{\ensuremath{p_{\mathrm{L}}}}
\providecommand{\NA}{\ensuremath{\text{---}}}

\cmsNoteHeader{B2G-17-003}
\title{Search for pair production of vector-like quarks in the $\bWbW$ channel from proton-proton collisions at $\sqrt{s} = 13\TeV$}

\date{\today}

\abstract{
A search is presented for the production of vector-like quark pairs,
\TTbar or \YYbar, with electric charge of
$2/3$ (\T)  or $-4/3$ (\Y), in proton-proton
collisions at $\sqrt{s} = 13\TeV$. The data were
collected by the CMS experiment at the LHC in 2016 and correspond
to an integrated luminosity of 35.8\fbinv.
The \T and \Y quarks are assumed to decay exclusively
to a \PW{}   boson and a \PQb quark. The search is based
on events with a single isolated electron or muon, large missing
transverse momentum, and at least four jets with large transverse momenta.
In the search, a kinematic reconstruction of the final state observables
is performed, which would permit a signal to be detected as a narrow mass peak
($\approx$7\% resolution). The observed number of events is consistent with
the standard model prediction. Assuming strong pair production of the
vector-like quarks and a 100\% branching fraction to \bW,
a lower limit of 1295\GeV
at 95\% confidence level is set on the \T and \Y quark
masses.
}

\hypersetup{
pdfauthor={CMS Collaboration},%
pdftitle={Search for pair production of vector-like quarks in the bWbW channel from proton-proton collisions at sqrt(s) = 13\TeV},%
pdfsubject={CMS},%
pdfkeywords={CMS, physics, vector-like  quarks}
}
\maketitle

\section{Introduction}
\label{sec:introduction}

Vector-like quarks (VLQs) are hypothetical spin-1/2 fermions, whose left-
and right-handed components transform in the same way under the standard
model (SM) symmetries, and hence have vector couplings. Nonchiral VLQs
appear in a number of beyond-the-SM scenarios, such as ``Randall--Sundrum"
and other extra-dimensional models~\cite{rs1,extradimensionsVLQ};
the beautiful mirrors~\cite{beautiful}, little
Higgs~\cite{littlesthiggs,Schmaltz200340,LittleHiggs1,littlehiggsreview,littleHiggs2},
and composite Higgs~\cite{compositehiggs} models;  grand unified
theories~\cite{GUTVLQ}; and also other models that provide insights into
the SM flavor structure~\cite{gaugeflavor}.
These models provide possible solutions to a number of problems, such as
electroweak symmetry breaking, a poor general fit to the precision
electroweak data, the origin of flavor patterns, and the hierarchy problem.
In particular, the hierarchy problem, namely the instability within the SM of
the Higgs boson mass parameter to quantum corrections, can be solved through
the introduction of VLQ contributions that cancel the contributions
from the top quark.

In general, VLQs T, with charge $+$2/3, are expected to mix significantly
only with the top quark, leading to the dominant \T quark decay
$\T\to \bW$~\cite{Aguila_VLQ_H}. We consider the case in which
this decay has a branching fraction
$\mathcal{B}(\T \to \bW) = 100\%$.
Also in some models~\cite{PhysRevD.88.094010}, a VLQ \Y with an electric
charge of $-$4/3 is predicted, either with or without the presence of
a \T VLQ. The \Y quark is expected to decay with a 100\% branching fraction
via the same \bW channel.
Since jets originating from the hadronization of \PQb quarks and \PAQb
antiquarks are not distinguished in this analysis, the results presented
apply equally to the strong pair production of both \T and \Y VLQs.
We consider the case where either only \T quarks or only \Y
quarks are produced.
This assumption produces a more conservative estimation of the lower mass
limit on VLQs. Throughout the rest of this paper, we will use \T to
represent both the \T and \Y VLQs.

In this paper, results are presented of a search for the strong pair
production of heavy VLQs and their subsequent decays through the signal
channel
\begin{equation*}
\TTbar \to \bW \PAQb \PW \to \PQb \ell \nu \PAQb \PQq \PAQq'
\end{equation*}
\noindent
in proton-proton (pp) collisions at $\sqrt{s} = 13$\TeV
using the CMS detector at the CERN LHC, where $\ell$ is an electron or
muon from the leptonic decay of one of the W bosons, and $\PQq$ and $\PAQq'$
are the quark and antiquark from the hadronic
decay of the other W boson. The analyzed data set
corresponds to an integrated luminosity of 35.8\fbinv.
This analysis is an extension to higher mass values of
an earlier CMS search for the \T quark at $\sqrt{s} = 8$\TeV.
Both the previous analysis and the present one are based on a kinematic
reconstruction with a constrained fit to the $\bWbW$ final state in the
signal decay channel shown above. Kinematic reconstruction enables detection
of the signal as a narrow mass peak.
The previous results
were combined with other CMS \T quark searches in Ref.~\cite{legacy_T23_CMS}.
The present observed lower mass limits for a \T quark decaying 100\% via the
bW channel are 920\GeV for CMS~\cite{legacy_T23_CMS} at $\sqrt{s} = 8$\TeV,
and 770\GeV at 8\TeV~\cite{T_bW_Atlas} and 1350\GeV at 13\TeV for
ATLAS~\cite{Atlas-13TeV-pap}.

The search strategy requires that one of the $\PW$ bosons decays leptonically,
producing an electron or a muon accompanied by a neutrino, and the other
decays hadronically to a quark-antiquark pair. We select events with a single
isolated muon or electron, missing transverse momentum, and at least
four jets with high transverse momenta, arising from the hadronization of
the quarks in the final state. We classify such events as $\mu$+jets  or
e+jets.

We perform a constrained kinematic fit on each selected event for the signal
decay process shown above. The full kinematic quantities of the final
state are reconstructed, and the invariant mass of the \T quark, $\Mreco$,
is obtained. We consider also cases when W bosons decaying hadronically at
high Lorentz boosts are reconstructed as single jets. Such merged jets are
then resolved into two subjets by employing jet substructure techniques
based on the``soft drop'' grooming algorithm~\cite{softdrop}. These resolved
subjets are counted individually when selecting four-jet final states and
contribute separately in the kinematic fit (see Section~\ref{sec:selection}).
Events with leptonically decaying W bosons include those decaying into a
$\tau$ lepton (in the decay sequence
$\PW\to \tau + \nu, \, \tau \to \ell + 2\nu)$. They are treated
in the same way as events with direct decays to muons or electrons.

\section{The CMS detector}
\label{sec:detector}

The central feature of the CMS apparatus is a superconducting solenoid of
6\unit{m} internal diameter, providing a magnetic field of 3.8\unit{T}.
Within the superconducting solenoid volume are a silicon pixel and strip
tracker, a lead tungstate crystal electromagnetic calorimeter (ECAL) with
preshower detector, and a brass and scintillator hadron calorimeter (HCAL),
each composed of a barrel and two endcap sections. Forward calorimeters
extend the pseudorapidity~\cite{CMS:2008zzk} coverage provided by the barrel
and endcap detectors. The detector is nearly hermetic, allowing for momentum
balance measurements in the plane transverse to the beam direction.
Muons are detected in gas-ionization chambers embedded in the steel
flux-return yoke outside the solenoid.

A more detailed description of the CMS detector, together with a definition
of the coordinate system used and the relevant kinematic variables, can be
found in Ref.~\cite{CMS:2008zzk}.

\section{Event samples}
\label{sec:data_mc}

The analysis is based on integrated luminosities of 35.8\fbinv
in the muon channel and 35.6\fbinv in the electron channel.
The trigger providing the muon data sample requires the presence
of at least one muon with $\pt > 50$\GeV and pseudorapidity
$\abs{\eta}<2.5$. For the electron data sample, events are required to have
a single isolated electron with $\pt > 32$\GeV and $\abs{\eta} < 2.1$.

Simulated event samples are used to estimate the signal efficiencies and
background contributions. The following background production processes
are modeled: $\ttbar$+jets; W+jets and Z+jets (single boson production);
single top quark via the tW, $s$- and $t$-channel processes; WW, WZ, and ZZ
(diboson production); and quantum chromodynamic (QCD) multijet production.
The dominant background is from $\ttbar$+jets production. All other background processes
are collectively referred to as non-$\ttbar$. The non-$\ttbar$ background
excluding multijets is called the electroweak background.

The \ttjets events are generated using the
\POWHEG v2.0~\cite{powheg1,powheg2,powheg3,Baglio:2015eon} event
generator. The diboson samples are produced using the
\PYTHIA 8.205~\cite{Sjostrand:8.2} generator.
The W+jets, Z+jets, and QCD multijet simulated events are produced with
the generator \MGvATNLO v2.2.2 \cite{MadGraph_2014}.
Single top quark events are generated with \POWHEG and
\MGvATNLO.

The simulated \TTbar signal events are generated with
\MGvATNLO at leading order for \T quark masses
from 800 to 1600\GeV in 100\GeV steps. The total $\TTbar$ inclusive
cross section ($\Pg\Pg \to \TTbar + X$) is computed for each \T quark
mass value at next-to-next-to-leading order, using a soft-gluon resummation
with next-to-next-to-leading-logarithmic accuracy~\cite{Czakon_Mitov}.
Signal samples are produced in the
narrow-width approximation in which the width of the generated \T quark
mass distribution of 1\GeV is much less than the mass resolution of
the detector.

The generated events are processed through the CMS detector simulation
based on \GEANTfour~\cite{geant4}.  Additional minimum-bias events,
generated with \PYTHIA, are superimposed on the hard-scattering
events to simulate multiple inelastic pp collisions in the same
or nearby beam crossings (pileup). The simulated events are weighted to
reproduce the distribution of the number of pileup interactions observed
in data, with an average of 23 collisions per beam crossing. All samples
have been generated with the NNPDF 3.0 set~\cite{NNPDF3} of parton
distribution functions (PDFs), using the tune CUETP8M1.
\PYTHIA is used to shower and hadronize all generated partons.

\section{Event reconstruction}
\label{sec:event_reco}

Events are reconstructed using a particle-flow (PF)
algorithm~\cite{particle-flow-pub} that combines information from all
the subdetectors: tracks in the silicon tracker and energy deposits
in the ECAL and HCAL, as well as signals in the preshower detector
and the muon system. This procedure categorizes all particles into
five types: muons, electrons, photons, and charged and neutral
hadrons. An energy calibration is performed separately for each
particle type.

Muon candidates are identified by multiple reconstruction algorithms
based on hits in the silicon tracker and signals in the muon system. The
standalone-muon algorithm uses only information from the muon chambers.
The silicon tracker muon algorithm starts from tracks found in the silicon
tracker and then associates them with matching signals in the muon detectors.
In the global-muon algorithm, for each standalone-muon track a matching
tracker muon is found by comparing parameters of the two tracks propagated
onto a common surface. A global-muon track is fitted by combining hits
from the silicon tracker muon and the standalone-muon track, using the
Kalman-filter technique~\cite{cms_muons}.
The PF algorithm uses global muons.

Electron candidates are reconstructed from clusters of energy deposited
in the ECAL matched with tracks in the silicon tracker. Electron tracks
are reconstructed using a dedicated modeling
of the electron energy loss and are fitted with a Gaussian sum filter
algorithm. Finally, electrons are further distinguished from charged
hadrons using a multivariate approach~\cite{Khachatryan:2015hwa}.

Charged leptons, originating from decays of heavy VLQs, are
expected to be isolated from nearby jets. Therefore, a relative isolation
parameter ($I_{\text{rel}}$) is used, which is defined as the sum of the \pt
of charged hadrons, neutral hadrons, and photons in a cone with distance
parameter $\Delta R=\sqrt{\smash[b]{(\Delta \phi)^2+(\Delta \eta)^2}}$ around the
lepton direction, where $\Delta\phi$ and $\Delta\eta$ are the azimuthal
and pseudorapidity differences, divided by the lepton \pt.
The isolation cone radius is taken as $\Delta R = 0.4$ for muons.
Pileup corrections to $I_\text{rel}$ are computed using tracks from
reconstructed vertices~\cite{cms_muons}.
For electrons, $\Delta R = 0.3$ and pileup corrections are calculated
using jet effective areas~\cite{fastjet1,fastjet2} separately for the
barrel and endcap regions.

Particles found by the PF algorithm are clustered into jets using the PF
jet identification procedure~\cite{particle-flow-pub}.
Using the charged-hadron subtraction (CHS) algorithm, charged hadrons
associated with pileup vertices are not considered, and particles that
are identified as isolated leptons are removed from the jet clustering
procedure. In the analysis, two types of jets are used: jets
reconstructed using the anti-\kt algorithm~\cite{antikt} with distance
parameters of either R = 0.4 (AK4) or 0.8 (AK8), as implemented in \FASTJET
v3.0.1~\cite{fastjet, hep-ph/0512210}. An event-by-event jet area-based
correction~\cite{Chatrchyan:2011ds,Khachatryan:2016kdb,fastjet1,fastjet2}
is applied to remove on a statistical basis pileup contributions that are
not already removed by the CHS procedure.

Since most jet constituents are identified and reconstructed with close to
a correct energy by the PF algorithm, only a small residual energy correction
must be applied to each jet. These corrections were obtained as a function
of jet \pt and $\eta$ from a comparison of \GEANTfour-based CMS Monte Carlo
(MC) simulation~\cite{cms-simulation} and collision data. Jet energy
corrections (JEC) are applied to each jet as a function of \pt and $\eta$.

For the identification of jets originating from the hadronization of a
$\PQb$ quark (\PQb-tagged jets) we use the combined secondary vertex
(CSVv2) algorithm~\cite{CMS-PAS-BTV-15-001}. This provides $\PQb$
tagging identification by combining information about impact parameter
significance, secondary vertex reconstruction, and jet kinematic
distributions. We use two operating points: medium (CSVM), with a mistagging
rate of 1\%, and loose (CSVL), with a 10\% mistagging rate, for which
the efficiencies of correctly tagging jets coming from \PQb quark hadronization
are 66\% and 82\%, respectively~\cite{CMS-PAS-BTV-15-001}.
The differences in the performance of the \PQb tagging algorithm in data
and MC simulation are accounted for by data/MC scale factors (SFs).

The missing transverse momentum in an event, \ptmiss, is defined as the
magnitude of the missing transverse momentum vector, which is the projection
on the plane perpendicular to the beams of the negative vector sum of the
momenta of all reconstructed particles in the event.
The vertex with the highest sum of squared \pt of all associated tracks
is taken as the hard-scattering primary vertex.

\section{Event selection and mass reconstruction}
\label{sec:selection}

Selected events are required to contain exactly one charged lepton (muon or
electron). Muon candidates are required to have $\pt > 55$\GeV and $\abs{\eta}< 2.4$.
The relative muon isolation parameter must satisfy $I_\text{rel} < 0.15$.
Selected electrons should have $\abs{\eta}< 2.1$. To ensure that the e+jets
channel covers a similar kinematic phase space to the $\mu$+jets channel,
electron candidates must satisfy the same $\pt > 55\GeV$ requirement.
Simulation shows that lowering the \pt threshold would not improve the signal
significance. Events with a second more loosely identified electron or muon,
with $\pt > 20 \GeV$ and $\abs{\eta} < 2.5$ (2.4) for electrons (muons), are
vetoed.

At the next step, we select a collection of jets that are used as input
to the kinematic fit. The collection includes AK4 jets that have  $\pt > $
30\GeV and $\abs{\eta} < 2.4$, and AK8 jets that satisfy $\pt > $  200\GeV
and $\abs{\eta} < 2.4$. A selected event must have either at least four AK4
jets, or at least three AK4 jets and at least one AK8 jet for the case where
the AK8 jet overlaps an AK4 jet (see explanations below). This is needed to
satisfy the requirement to have at least four jets in the final jet
collection used for the kinematic fit.

As the mass of a heavy VLQ increases, the reconstructed topology of its
decay is modified by the overlapping and merging of jets owing to the high
Lorentz boosts its decay products receive. The quark pair from the hadronically
decaying W boson becomes increasingly collimated, producing overlapping
hadronic showers that cannot be resolved as separate jets. This precludes
performing constrained kinematic fits that use jets as proxies for the
final-state quarks in the signal decay process.

The AK8 jets are used to identify the merged hadronic W boson decays by
applying the ``soft drop'' (SD) grooming algorithm~\cite{softdrop}, with
parameters $z_\text{cut} = 0.1$ and $\beta = 0$. This procedure removes
soft and wide-angle radiation from jets and resolves the internal structure
of the wide jet into two distinct subjets, associated with the underlying
W boson decay. The JECs that are applied to the AK8 jets are propagated to
the pair of subjets by scaling them so that the sum of their four-momenta
is equal to the parent AK8 jet four-momentum. The groomed jet mass, called the soft drop mass,
$m_\mathrm{SD}$, is taken as the invariant mass of the constituents of the
AK8 jet with the SD algorithm applied, and thus is equal to the invariant
mass of the two subjets. It is required to be within the W boson mass window
$60 < m_\mathrm{SD} <100 \GeV$, in which case the AK8 jet is labeled
``W-tagged''.

For each W-tagged AK8 jet, representing an identified footprint of a highly
boosted, hadronically decaying W boson, we search for the corresponding
footprint of the same W boson decay in the AK4 jet collection.
This is done by trying to match a W-tagged AK8 jet to an AK4 jet or to
the vector sum of a pair of nearby AK4 jets for which
$\DR \mathrm{(AK8,\, AK4)} < 0.8$. Since the opening angle of AK8 subjets,
resolved by the SD algorithm, can go down up to $\DR = 0.15$, the system
of the two subjets represents a more accurate assignment of jet constituents
to two separate subjets compared to the clustering into two AK4 jets, with
their more coarse threshold of $\DR = 0.4$.
A match requires $\DR{(AK8,\, AK4)} = \DR_{\text{match}} < 0.05$.
Matching with a pair of AK4 jets is preferred to a matching with only one
AK4 jet, when the $\DR_{\text{match}}$ value for that pair is smaller than the
values for any of the single jets in the pair. In this case matching with
the pair (represented by the vector sum of the two AK4 momenta) provides
the best association of AK4/AK8 jets.

A matched AK4 jet is replaced by the two subjets of the W-tagged AK8
jet, thus using the full kinematic information obtained by the W tagging
and including that into a jet collection used for the kinematic fit. This
procedure results in a hybrid jet collection, consisting of AK4 jets left
unmatched and subjets of W-tagged AK8 jets, thus excluding the possibility
of double counting the jets later used as input to the kinematic fit.

Only the jets in the hybrid collection are used in the rest of the analysis.
A set of preselection requirements is now applied to the event:
\begin{itemize}
\item
The missing transverse momentum in the event must satisfy
$\ptmiss > 30 \GeV$. This requirement is designed to both select final
states containing neutrino and suppress QCD multijet background contribution.
\item
Each of the jets must have  $\pt >  30$\GeV and $\abs{\eta} < 2.4$ (these requirements
were not applied to the newly included W-tagged subjets).
\item
Jets too close to the lepton direction are discarded by requiring $\DR (\text{jet},\, \ell) > 0.4$.
\item
There must be at least 4 remaining jets in the event after the previous
criteria.
\item
The two highest-\pt jets must satisfy $\pt > 100$ and 70 \GeV,
respectively.
\end{itemize}

Events that pass all these requirements are used as input to the kinematic
fit, which is performed using the \textsc{HitFit} package~\cite{hitfit_web}.
This fitting program was developed by the D0 experiment~\cite{snyder}
for the measurement of the top quark mass in the lepton+jets channel.
It contains a fitting engine, which minimizes a  $\chi^2$ quantity subject
to a set of constraints, and an interface.

The input to the fitting engine comprises the two-dimensional vector \ptvecmiss
and the measured three-dimensional momenta of the charged lepton and four jets.
The four quarks in the final state of the signal decay process manifest
themselves as jets whose measured three-momenta are used as estimates of
the quark momenta. The \ptmiss in the event is used as an estimate of the
transverse momentum of the neutrino. The unmeasured longitudinal component
of the neutrino momentum is calculated from one of the kinematic constraints
shown in Eqs.~(\ref{eq:con1}) and (\ref{eq:con3}) below.

The fit is performed by minimizing a $\chi^2$ computed from
the differences between the measured momentum components and their fitted
values, divided by the corresponding uncertainties, summed over all the
reconstructed objects in the final state. The fit is subject to the following
constraints:
\begin{align}
\label{eq:con1}
m(\ell\Pgn)&= m_{\PW}, \\
\label{eq:con2}
m(\cPq\cPaq')&= m_{\PW}, \\
\label{eq:con3}
m(\ell \Pgn \cPqb) &= m(\cPq\cPaq'\cPqb) = m_{\text{reco}},
\end{align}
where $m_{\PW}$ is the $\PW$ boson mass~\cite{PDG2016}, $\ell$ stands for
electron or muon, and  $\ell$, $\Pgn$, $\cPqb$ can denote either a particle
or antiparticle. Equation~(\ref{eq:con2}) requires that the invariant mass
of the quark-antiquark pair equals the W boson mass. Equation~(\ref{eq:con3})
demands that the reconstructed invariant masses $\Mreco$ of
the two produced \T quarks are equal. After one constraint is used for the
calculation of the longitudinal neutrino momentum, two constraints are left.
To check to what extent the assumed kinematic hypothesis is
compatible with the fitted momenta, a so-called ``goodness-of-fit'' is
calculated, given by the probability $P(\chi^2 \ge \chi^2_\text{min})$ for the
$\chi^2_\text{min}$ value obtained after minimization.

For events with exactly 4 jets, all jet permutations
in which jets are assigned specific quark roles in the final state of the
signal process are prepared by the \textsc{HitFit} interface
and entered in turn into the fitting engine. If there are more than four jets
in an event passing the jet requirements, then the five jets with the highest
\pt are considered and all permutations of four jets out of five are used
as input to the fitter. The calculation of the longitudinal component of
the neutrino momentum has a two-fold ambiguity from solving a quadratic
equation. This doubles the number of fitted combinations, based on the jet
permutations. At the end of the fitting process, \textsc{HitFit} delivers
information about 24\,(120) fitted permutations for the case of 4\,(5) jets.

After the event fitting is done, we have to decide which fitted combination
must be chosen to represent the final state of the signal process.
To reduce the number of accidental combinatoric assignments, we use information
on \PQb tagging and W tagging. Pairs of \PW{}-tagged subjets (if available) are
assigned to the hadronic W boson decay.
To identify the most likely lepton+4 jets combination arising from the
decay chain of the signal process, we inspect each fitted combination
in turn, proceeding as follows:

\begin{itemize}
\item
In each fit combination, two jets are taken to be the \PQb jets from the
$\T$ and $\Tbar$ decays. We call $\mathrm{b_l}$ the \PQb jet accompanying the
leptonic W boson decay, and $\mathrm{b_h}$ the \PQb jet accompanying the hadronic
W boson decay.  Combinations in which neither of these jets is b-tagged or
only one has a CSVL tag are rejected.
\item
The four jets in the fitted combination, designated in the order:
$\mathrm{b_h}$ jet, $\mathrm{b_l}$ jet, highest-\pt jet in the hadronic W
boson decay, second-highest-\pt jet in the hadronic W boson decay, ~must
satisfy the requirements
$\pt > 200,  100, 100$, and 30\GeV, respectively.
\item
For each fitted jet combination a variable $S_\T$ is calculated, defined
as the scalar sum of \ptmiss and the transverse momenta of the lepton and
the four jets in that combination:
$S_\Trm= \ptmiss+\pt^{\ell}+\pt^{J_{1}}+\pt^{J_{2}}+\pt^{J_{3}}+\pt^{J_{4}}$,
where $J_i$ (with $i = $ 1 to 4) refers to the four jets and $S_\T$ is evaluated
using the measured momenta. To select hard-scattering processes resulting in
the production of heavy objects, we require $S_\Trm > 1000 \GeV$.
\item
 $S_{\Lrm}^{\text{fit}} / S_{\Trm}^{\text{fit}} <  1.5$, where
$S_{\Lrm}^{\text{fit}}= \pL^{\nu}+\pL^{\ell}+\pL^{J_{1}}+\pL^{J_{2}}+\pL^{J_{3}}+\pL^{J_{4}}$,
and $\pL$ is the longitudinal momentum of each of the corresponding objects.
Both $S_{\Lrm}^{\text{fit}}$ and $S_{\Trm}^{\text{fit}}$
are calculated using the fitted momenta.
This requirement relies on the fact that the final-state objects from the
signal process typically have both high-\pt and moderate-$p_z$ values.
\item
The invariant mass of the two jets attributed to the W boson hadronic decay
must be in the range 60--100\GeV.
\item
$P(\chi^2) > 0.1\%$ (corresponding to 3 standard deviations for a one-sided Gaussian distribution).
\end{itemize}

The signal decay process contains two \PQb quarks,
which manifest themselves as b-tagged jets. The fit combinations passing the
above selection are sorted into four groups according to the \PQb tagging categories
of the \{$\mathrm{b_l, b_h}$\} jet pair: two CSVM \PQb tags; one CSVM and one
CSVL \PQb tag; only one CSVM \PQb tag; and two CSVL \PQb tags. The combination with the
largest $P(\chi^2)$
value from the group with the tightest \PQb tagging is selected for the signal
sample.

Combinations containing W-tagged subjets are considered first, if such
exist in the event. If no combination with W-tagged subjets passes the
selection criteria, then the other combinations are considered.

\subsection{Results of the event selection}

Table~\ref{tab:events} presents the numbers of observed and predicted
background events,
normalized to the integrated luminosity.
Selection efficiencies for the $\TTbar$ signal, including acceptance and
branching fractions of decays, and the number of expected signal events
are given in Table~\ref{tab:SignalEff} as a function of the \T quark mass.

\begin{table*}[htbp]
\topcaption{
The numbers of expected background events for each process in the $\mu$+jets
and e+jets channels, normalized to the integrated luminosity of the data,
the number of observed events, and the ratio of the observed to predicted
events. The uncertainties in the predicted numbers of background events are
statistical only.
}
\newcolumntype{x}{D{,}{\,\pm\,}{3.3}}
\label{tab:events}
\centering
\begin{tabular}{lxx}
                     & \multicolumn{1}{c}{$\mu$ + jets}  & \multicolumn{1}{c}{e + jets} \\
\hline
Background process   &                &  \\
\ttbar+jets &   533,6         &   470,5\\
Single top    &   115,5         &   100,4\\
W+jets        &  94,2       &  73,2 \\
Z+jets        &   10.7,0.3   & 9.4,0.3  \\
QCD multijet  &   8.8,4.4   &  15,8  \\
Diboson       &   4.4,4.4   &  1.8,1.8  \\
\ttbar V     &   10.9,0.9    &  8.0,0.8  \\[\cmsTabSkip]
Total background (MC)&   777,10    &  678,11  \\[\cmsTabSkip]
Total observed (Data) &   \multicolumn{1}{c}{768}    & \multicolumn{1}{c}{684}  \\[\cmsTabSkip]
Data/MC       &    0.99,0.04    &  1.01,0.04 \\
\end{tabular}

\end{table*}

\begin{table*}[htbp]
\topcaption{
Selection efficiencies from MC simulation for the $\TTbar$ signal as
a function of the \T quark mass assuming
$\mathcal{B}(\T \to \bW) = 100\%$, and the numbers
of expected signal events after the final selection for the integrated
luminosity of the data. The uncertainties are statistical only.
}
\label{tab:SignalEff}
\newcolumntype{x}{D{,}{\,\pm\,}{3.3}}
\centering
\begin{tabular}{cxx{c}@{\hspace*{5pt}}xx}
            & \multicolumn{2}{c}{$\mu$ + jets} && \multicolumn{2}{c}{e + jets} \\ \cline{2-3} \cline{5-6}
Mass [\GeVns{}] & \multicolumn{1}{c}{Efficiency (\%)} & \multicolumn{1}{c}{Events}          & &  \multicolumn{1}{c}{Efficiency (\%)}   & \multicolumn{1}{c}{Events} \\
\hline
800   & 2.37,0.05 &  166.1,3.7   &&  2.20,0.05 &  153.5,3.5  \\
900   & 2.71,0.06 &  87.8,1.9     &&  2.36,0.05 &  75.8,1.8   \\
1000  & 2.97,0.06 &  46.7,0.9    &&  2.69,0.06 &  42.2,0.9   \\
1100  & 3.08,0.06 &   24.7,0.5   &&  2.77,0.06 &  22.1,0.4   \\
1200  & 3.14,0.06 &  13.3,0.3    &&  2.80,0.06 &  11.8,0.2   \\
1300  & 3.10,0.07 &  7.09,0.15  && 2.79,0.06 &  6.34,0.14   \\
1400  & 3.14,0.06 &  3.98,0.08  && 2.77,0.06 &  3.50,0.07   \\
1500  & 3.21,0.06 &  2.30,0.04  && 2.76,0.06 &  1.96,0.04   \\
1600  & 3.01,0.06 &  1.24,0.02  && 2.84,0.06 &  1.16,0.02   \\
\end{tabular}

\end{table*}

\begin{figure}[htbp]
  \centering
  \subfigure{\includegraphics[width=\cmsFigWidth]{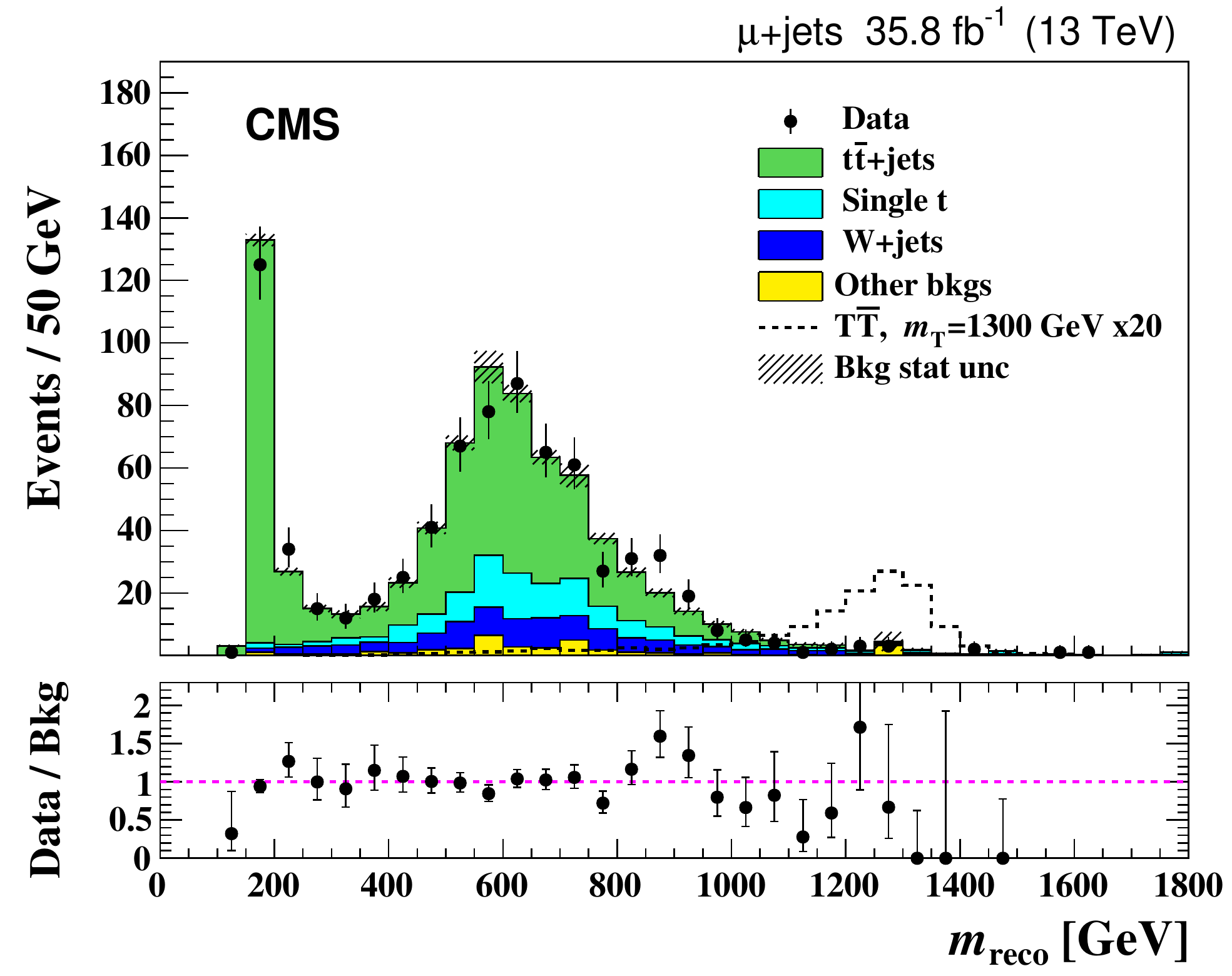}}\\
  \subfigure{\includegraphics[width=\cmsFigWidth]{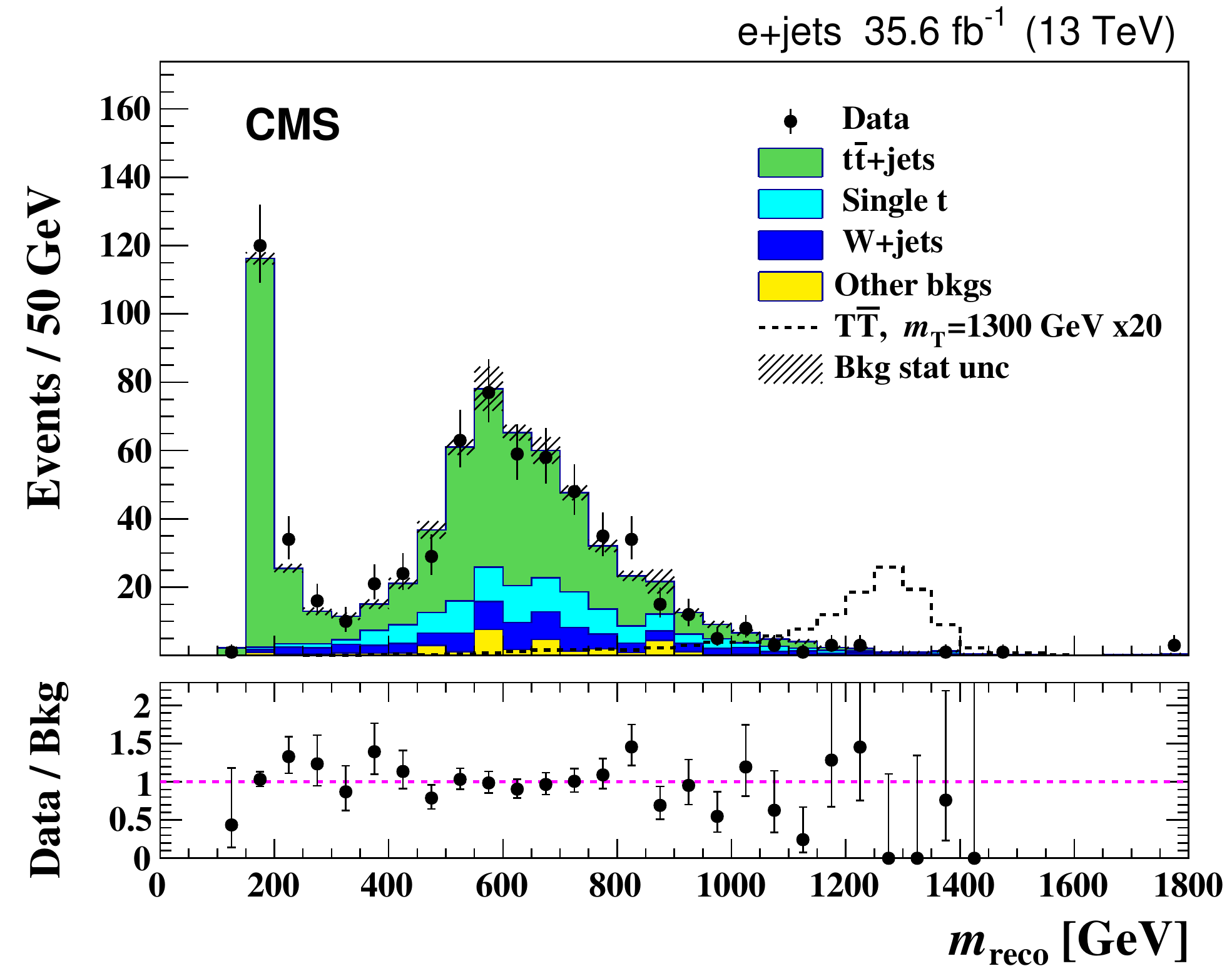}}\\
  \caption{
The \T quark reconstructed mass spectra for the $\mu$+jets (upper plot) and
e+jets (lower plot) channels from data and from MC simulations of
signal and background processes.
The MC prediction for pair production of a \T quark
with a mass of 1300\GeV is shown by a dashed line, enhanced by a factor of 20.
The lower panels show the ratio of the data to the background
prediction. The uncertainties represented by the vertical bars on the points
are statistical only. The shaded regions show the total statistical uncertainties
in the background.
}
  \label{fig:mfit_l}
\end{figure}

\begin{figure}[htbp]
  \centering
  \subfigure{\includegraphics[width=\cmsFigWidth]{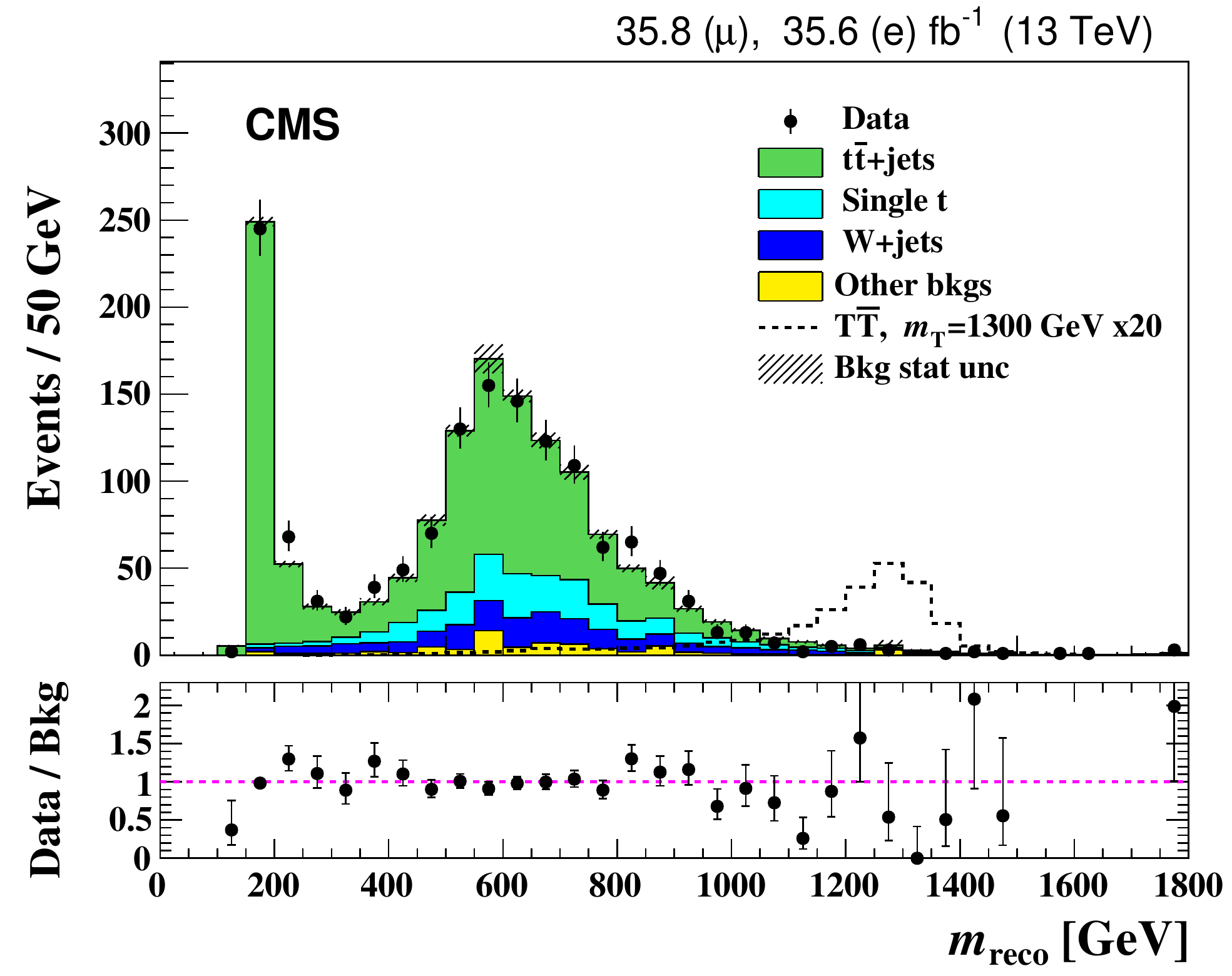}}\\
  \subfigure{\includegraphics[width=\cmsFigWidth]{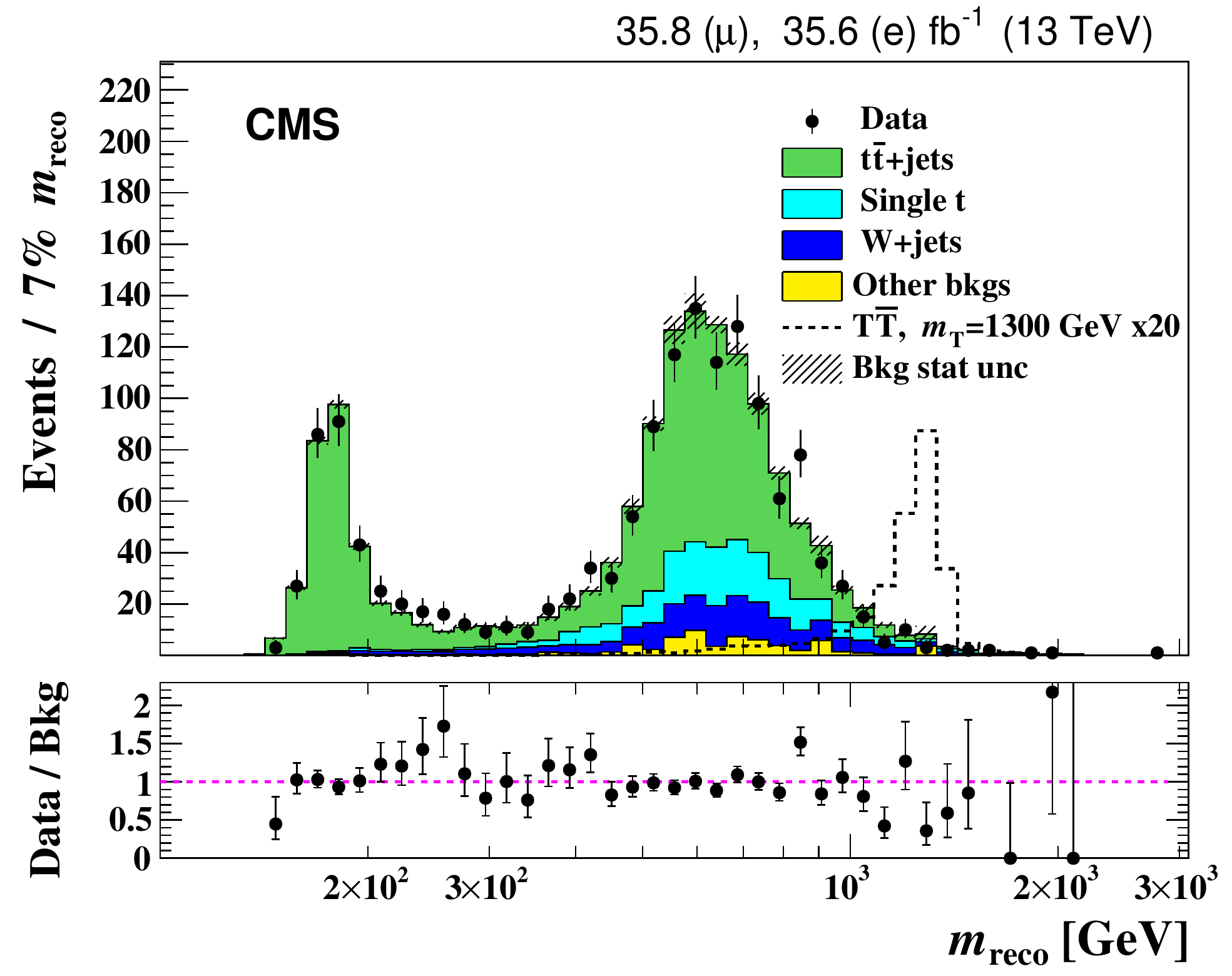}}
  \caption{
The \T quark reconstructed mass spectra for the sum of $\mu$+jets and e+jets
channels from data and from MC simulations of signal and background
processes. The lower plot has a logarithmic x-axis scale and
the bin size corresponds to 7\% of the mass value in the middle of the bin.
The MC prediction for the pair production of a \T quark with a mass of 1300\GeV
is shown by a dashed line, enhanced by a factor of 20.
The lower panels show the ratio of the data to the background
prediction. The uncertainties represented by the vertical bars on the points
are statistical only. The shaded regions show the total statistical uncertainties
in the background.
}
  \label{fig:mfit_sum}
\end{figure}

Figure~\ref{fig:mfit_l} shows the distributions of the \T quark reconstructed
mass, $\Mreco$, obtained from the selected $\mu$+jets and e+jets events,
and Fig.~\ref{fig:mfit_sum} shows their sum. The integrated luminosity and
cross sections of the background processes are used for normalization of
the background contributions. Only the statistical uncertainties are shown
for the estimated background, since the systematic uncertainties enter
the workflow later and are not available at this point of the analysis.
Nevertheless, it is evident that the data distributions are well described
by the predicted backgrounds alone.

The resolution of the reconstructed mass was found to be proportional to
the value of the mass, with the Gaussian core of the  $\Mreco$ resolution
curves having a width-to-mass ratio of $\approx$7\%. We make use of this
feature by employing variable bin widths equal to 7\% of the bin-center
values when plotting  $\Mreco$ and using a logarithmic scale for the
horizontal axis. In this way, the bin widths appear to be of the same size,
equal to the local mass resolution.
This helps smooth statistical fluctuations causing peaks that are narrower
than the local mass resolution, and avoid wrong interpretations when
searching for narrow structures in the observed mass distribution.
Figure~\ref{fig:mfit_sum} (lower plot) shows the resulting $\Mreco$
spectrum plotted this way for the sum of the $\mu$+jets and e+jets channels.
The reconstructed top quark peak from the $\ttbar$+jets background process
and the predicted signal for a \T quark with a mass of 1300\GeV, enhanced
by a factor of 20 for better visibility, appear in the figure as narrow
peaks with similar widths.

The analysis selection has been optimized to obtain the best signal
significance at high masses, and the mass reconstruction in the region
of the top quark has not received special attention.
As a result, the bin width used in Fig.~\ref{fig:mfit_l} and the upper
part of Fig.~\ref{fig:mfit_sum} is too coarse to reveal details of the
mass reconstruction in this region. Nonetheless, the top quark peak
provides a useful benchmark for checking the reconstructed mass
scale and resolution in data and Monte Carlo.

\section{Systematic uncertainties}
\label{sec:systematics}

In this section, we describe the systematic uncertainties in the
calculation of the signal cross section. The uncertainties can be divided
into two categories: those that only impact the normalizations of the
distributions, and those that also affect the shapes of the distributions.
Each systematic uncertainty is included as a nuisance parameter in
the likelihood fit described in Section~\ref{sec:limits}.

The uncertainties in the $\ttbar$+jets, electroweak, and QCD multijet
cross sections, the total integrated luminosity, and the lepton
efficiencies affect only the normalization.

The uncertainty in the cross section for $\ttbar$+jets production of 5.3\%
is taken from a previous CMS measurement~\cite{CMS-ttbar-13TeV}.
The integrated luminosity is known to a precision of 2.5\%~\cite{lumi2016}.
A 10\% uncertainty is assigned to the sum of the non-\ttbar
backgrounds, which is dominated by the uncertainty in the W+jets and single
top quark backgrounds. This uncertainty is obtained from earlier CMS
measurements on single top quark production~\cite{cms-single-top}
and from a preliminary CMS measurement on inclusive W production.

Trigger and lepton identification efficiencies and data/MC SFs
are obtained from data using decays of \cPZ\ bosons to
dileptons. The systematic uncertainty in the lepton identification and
trigger efficiencies is 2.5\%.

Uncertainties that affect the shapes of the $\Mreco$ distributions include
those in the jet energy scales (JES), jet energy resolution (JER), $\PQb$
tagging efficiency, pileup, renormalization/factorization scale, and PDFs.
To model these uncertainties, we produce additional distributions, called
templates, by varying the nuisance parameter that characterizes each
systematic effect by one standard deviation up and down. To determine the
signal and background templates for any value of the nuisance parameter, we
interpolate the content of each bin between the varied and nominal templates.
This procedure is often referred to as vertical morphing~\cite{morphing}.

The JES uncertainty affects the normalization and the shape
of the $\Mreco$ distribution. This is taken into account by generating $\Mreco$
distributions for values of the jet energy scaled by one standard deviation
of the $\eta$- and \pt-dependent uncertainties.

The MC was found to underestimate the JER observed in the data, and as
a result the MC simulated jets are smeared to describe the data.
To this end, the difference between the reconstructed jet \pt and the generated
jet \pt is scaled by $\eta$-dependent SFs (the so-called ``nominal''
variation). To estimate the  uncertainty in this rescaling, the analysis
is repeated twice, each time applying additional sets of SFs that
correspond to varying the nominal ones up and down by one standard deviation.
Both  AK4 and AK8 jets are subject to JER systematic variations.
The JES and JER systematic variations are applied before the AK8 jet splitting.
Systematic variations of each subjet are done with the same relative variation
as the entire AK8 jet, so that their sum is equal to the modified AK8
four-momentum. The resulting jet momentum changes in the AK4 jet collection
are propagated to \ptmiss.

The systematic uncertainty related to the $\PQb$ tagging efficiency is
estimated by varying the \PQb tagging SFs for both the medium and loose operating
points by one standard deviation separately for heavy-flavor and light-flavor jets .

The pileup uncertainty is evaluated by varying the minimum-bias cross section
used to calculate the pileup distribution in data by $\pm$4.6\%, and
adjusting the number of pileup interactions in the simulation to these
distributions. This variation is taken from the uncertainty in a preliminary
CMS measurement of the minimum-bias cross section.

The uncertainties due to variations in the renormalization and factorization
scales are evaluated using per-event weights, corresponding to
renormalization/factorization scale variations by a factor of two up and down.
The combinations of scales corresponding to unphysical anti-correlated
variations are not considered. The envelope of the observed
variations in the $\Mreco$ spectrum is taken as an estimate of the uncertainty.

For evaluating the uncertainty related to the PDFs, we use 100 MC replicas,
generated with the NNPDF3.0 PDF set and their corresponding weights, to
sample the $\Mreco$ distribution. The per-bin RMS in the $\Mreco$ distribution
is taken as a measure of the PDF uncertainty in the corresponding templates.
The PDF uncertainty is applied both to the background and the signal MC
replicas. In the case of the signal, the uncertainty affects both the shape
and yield, though the latter is only due to the signal acceptance, not to
the signal cross section.

It is known that the top quark \pt distribution in the $\ttbar$+jets background
process is not well modeled by the MC simulation~\cite{pt-reweigh}.
Therefore, a reweighting of the top quark MC \pt distribution is applied. An
event weight is calculated, based on the generator-level top
quark \pt. The systematic uncertainty related to the top quark \pt distribution
reweighting is determined from the difference between applying and not applying the
reweighting,
and by applying the reweighting twice.

For the W+jets background modeling, we use \HT-binned MC samples, for
which the generator-level \HT distribution was found to deviate from the
same distribution in the inclusive W+jets simulation. The variable \HT is
defined as the scalar sum of the transverse momenta of all partons
in a simulated event that originate from the hard-scattering process.
In each \HT-binned sample, events are simulated in a certain range of \HT
(200 to 400\GeV, 400 to 600\GeV, etc.).
To improve the modeling, we implement an event-weighting technique in
which each simulated W+jets event is weighted depending on its \HT value.
The weight is based on a parametrization obtained from a fit to the ratio
of the generator-level \HT distributions of inclusive and \HT-binned
samples. The systematic uncertainty related to the event weighting is
estimated from the difference between applying and not applying
the weighting,
and by applying the weighting twice.

To estimate the systematic uncertainties related to the $\DR_{\text{match}}$
requirement used to associate W-tagged AK8 jets with AK4 jets in the
jet-splitting procedure, two templates with $\pm$20\% variation in the
maximum allowed $\DR_{\text{match}}$ value were prepared.

Table~\ref{tab:uncerts} shows the influence of the shape systematic
uncertainties on the signal, for a \T mass of  1200\GeV, and background
yields. Other sources of systematic uncertainties have a negligible impact
on the analysis.

\begin{table*}[htbp]
\topcaption{
Variations in percent on the yield of the selected MC events due to shape systematic
uncertainties for a signal with a \T quark mass of 1200\GeV and background.
}
\label{tab:uncerts}
\centering
\begin{tabular}{lcc}
                    & Signal (\%) & Background (\%) \\
\hline
JES                       & $+$0.2, $-$2.5  & 17 \\
JER                       & $+$0.02, $-$0.3 & 0.03 \\
b tag heavy flavor        & 2.5             & $+$2.8, $-$1.4 \\
b tag light flavor        & 0.2             & 0.8 \\
Renorm./fact. scales      & 1.1             & $+$18, $-$14 \\
Pileup                    & 0.05            & 0.2 \\
PDF                       & 0.3             & 2.0\\
Top quark \pt reweighting & \NA               & 11 \\
W+jets reweighting        & \NA               & $+$4.9, $-$3.3 \\
$\DR_{\text{match}}$       & $+$0, $-$0.8    & $+$0, $-$1.9 \\
\end{tabular}
\end{table*}

\section{Cross section and mass limits}
\label{sec:limits}

Since the observed distributions are consistent with the expected background,
we use the results to set limits on the $\TTbar$ production cross section
and on the \T quark mass.

\begin{figure}[htbp]
\centering
\includegraphics[width=\cmsFigWidth]{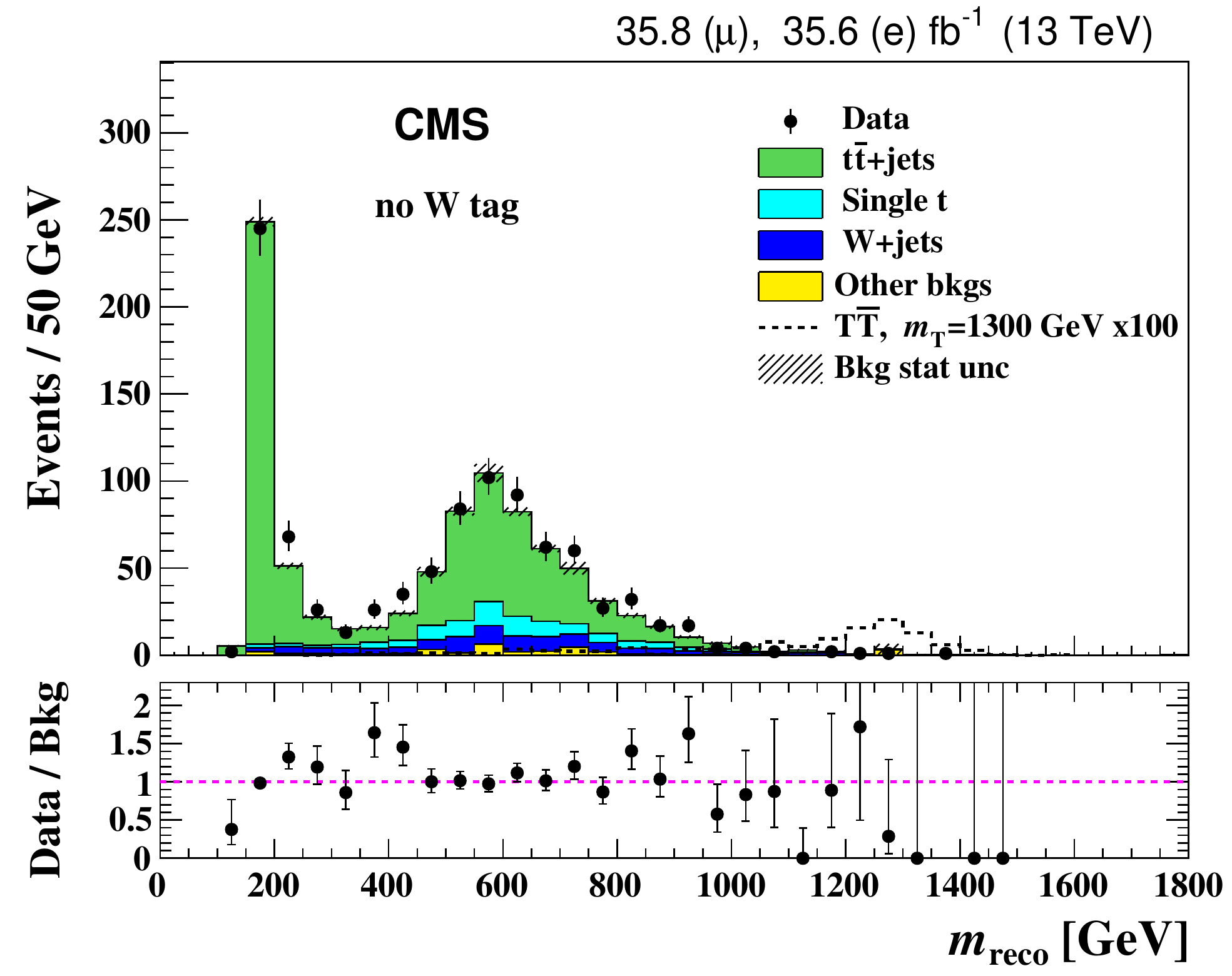}\\
\includegraphics[width=\cmsFigWidth]{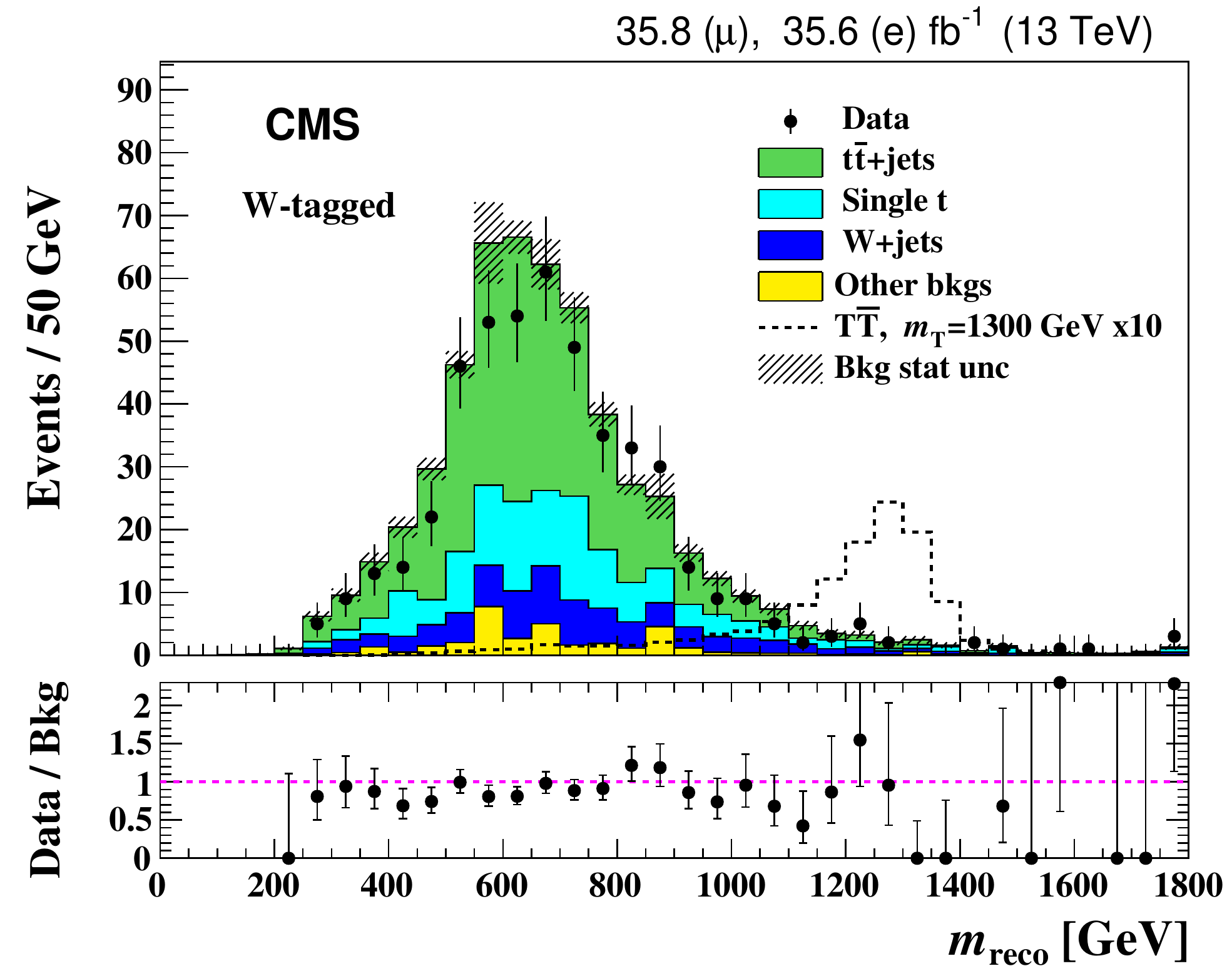}\\

\caption{The \T quark reconstructed mass spectra for the sum of the $\mu$+jets
and e+jets channels for the ``no W tag'' category of events (upper plot), and for
the ``W-tagged'' category (lower plot) from data and from MC simulations
of the background processes. The MC prediction
for $\TTbar$ production of a \T quark with a mass of 1300\GeV is shown by
a dashed line, enhanced by a factor of 100 (upper) and 10 (lower).
The lower panels show the ratio of the data to the background
prediction. The uncertainties represented by the vertical bars on the data
points are statistical only.
}
\label{fig:Mfit_categories}
\end{figure}

As discussed in Section~\ref{sec:selection}, there are two types of selected
events: those with W-tagged jets (labeled as ``W-tagged''), and those without
(labeled as ``no W tag''). The $\Mreco$ invariant mass distributions for these
two categories of events are shown separately in Fig.~\ref{fig:Mfit_categories}
for data and simulated background, combining the $\mu$+jets and e+jets
contributions. The no W tag category is more sensitive to lower-mass $\TTbar$
signal events, as shown in the upper plot of Fig.~\ref{fig:Mfit_categories}
by the clear top quark mass peak. At high masses this category of events gives
a very small contribution to the signal. Conversely, the W-tagged category is more
sensitive to high-mass signal events, as can be seen in
Fig.~\ref{fig:Mfit_categories} from the predicted distributions for a $\TTbar$
signal with a \T mass of 1300\GeV for the two categories. We therefore use the data
$\Mreco$ distribution from the W-tagged events in setting a lower limit on
the \T quark mass. We checked that using the no W tag category together with
the W-tagged category did not improve the sensitivity at high masses.

\begin{figure}[htbp]
\centering
 \includegraphics[width=\cmsFigWidth]{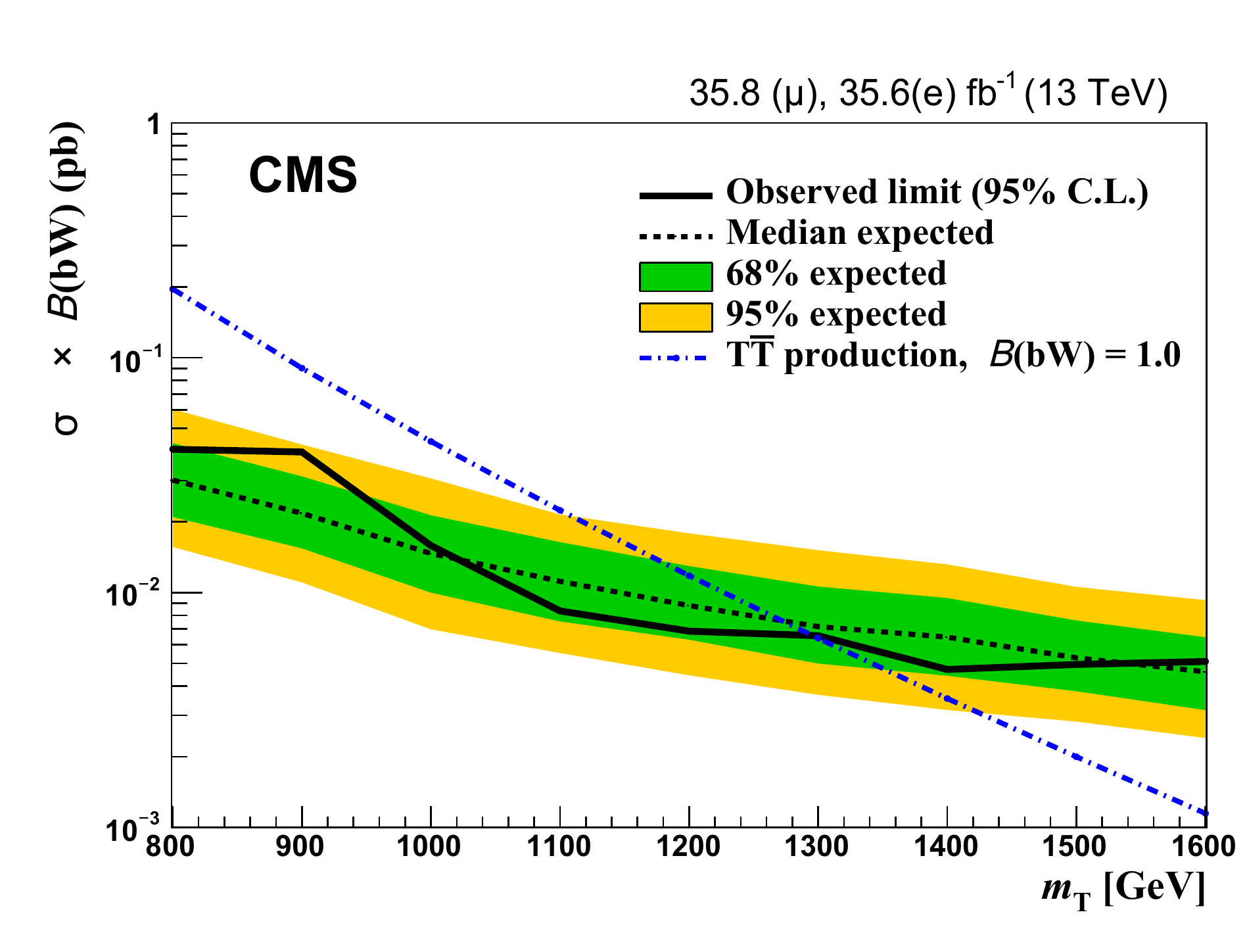}
 \includegraphics[width=\cmsFigWidth]{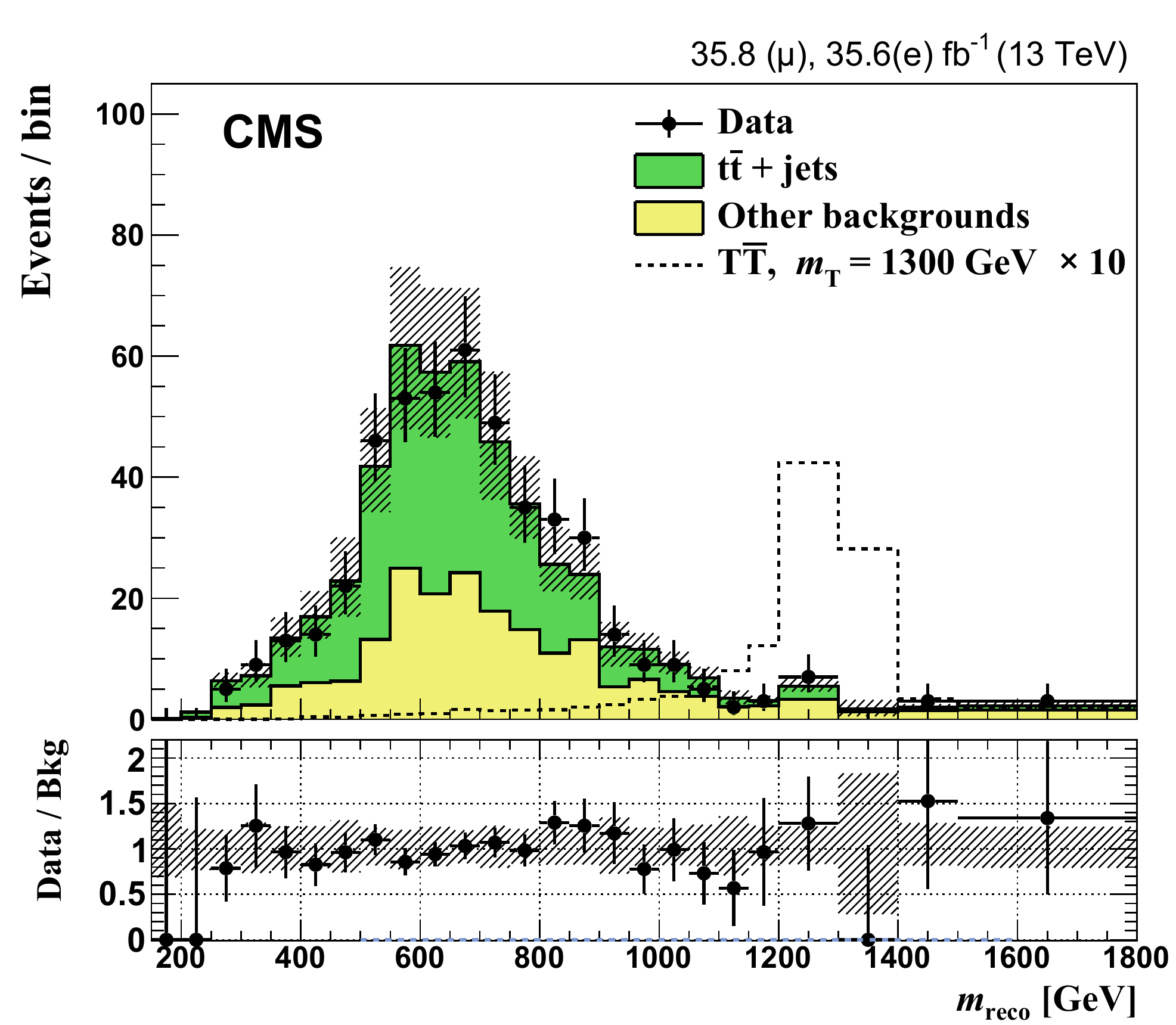}
\caption{Upper plot: Observed and expected Bayesian upper limits at 95\% CL on the
product of the $\TTbar$ or $\YYbar$ production cross section and the branching
fraction to \bW using only the W-tagged events. The inner and outer bands
show the 1 and 2 standard deviation uncertainty ranges in the expected limits,
respectively. The dashed-dotted line shows the prediction of the theory.
Lower plot: The post-fit distribution of the reconstructed \T quark mass, $\Mreco$.
Horizontal bars on the data points show bin size.
The shaded band on the histogram and on the ratio plot shows the quadrature sum
of the statistical and systematic uncertainties. The MC prediction for heavy quark
production
with a mass of 1300\GeV is shown by a dashed line, enhanced by a factor of 10.
}
\label{fig:Limits_Wtag}
\end{figure}

The distribution shown in Fig.~\ref{fig:Mfit_categories} (lower plot) is used for
the limit calculations. For the MC background distribution we require that the
relative statistical uncertainty in each bin is not worse than 20\%. For masses
above 1200 GeV the bins are merged to meet this requirement. The final binning
can be seen on the post-fit distribution shown in Fig.~\ref{fig:Limits_Wtag}
(lower plot).

The 95\% confidence level (CL) upper limits on the production cross section of
$\TTbar$ are computed within the \textsc{Theta} framework~\cite{Theta_web} using
a Bayesian interpretation~\cite{bayesian}, in which the systematic uncertainties
are taken into account as nuisance parameters.
The binned maximum-likelihood fit to the data distribution
is performed with a combination of the background contributions plus a signal.
The main backgrounds are $\ttbar$+jets, single top quark, and W+jets production.
Other smaller backgrounds, including electroweak and QCD multijet processes,
are summed up with the single top quark and W+jets contributions and are called
non-$\ttbar$ background (see Table~\ref{tab:events}). Distributions of
possible $\TTbar$ signals are considered for \T masses from 800 to 1600\GeV
(Table~\ref{tab:SignalEff}).

The likelihood function is marginalized with respect to the nuisance parameters
representing the systematic uncertainties in the shape and normalization.
Thirteen nuisance parameters are employed in the likelihood fit: $\ttbar$+jets
cross section uncertainty of 5.3\%, normalization of the non-$\ttbar$
contribution of 10\%, the integrated luminosity uncertainty of 2.5\%, lepton
identification and trigger uncertainty of 2.5\%; other uncertainties are the
shape uncertainties that include the JES, JER,
the \PQb tag SFs for light- and heavy-flavor jets, the renormalization and
factorization scales, pileup, PDFs, top quark \pt reweighting, and W+jets
background reweighting. Shapes of the SM background and signal templates are
changed (``morphed") in the limit-setting procedure according to the varying
values of the shape nuisance parameters. Contributions from the SM processes
are allowed to vary independently within their systematic uncertainties,
using log-normal priors~\cite{log_prior_ref1,log_prior_ref2}. The nuisance
parameters describing the shape uncertainties are constrained using Gaussian
priors. A flat prior probability density function on the total signal yield
is assumed. The MC statistical uncertainties in the simulated samples are
also included in this calculation.

Figure~\ref{fig:Limits_Wtag} (upper plot) shows the observed and expected
95\% CL upper limits on the $\TTbar$ cross section as a function of the
T quark mass.
Figure~\ref{fig:Limits_Wtag} (lower plot) shows the post-fit distribution
of the reconstructed mass. From the upper cross section limits we set lower
limits on the \T quark mass. The 95\% CL lower limit on
the \T mass is given by the value at which the 95\% CL upper limit curve
for the $\TTbar$ cross section intersects the theory curve, as shown in
Fig.~\ref{fig:Limits_Wtag}. In the $\bWbW$ channel with an assumed
branching fraction $\mathcal{B} (\T \to \bW)$ = 100\%, the observed
(expected) lower limit on the \T quark mass is 1295\,(1275)\GeV.

\section{Summary}
\label{sec:summary}

The results of a search for vector-like quarks, either $\T$ or $\Y$, with
electric charge of 2/3 and $-$4/3, respectively, that are pair produced in
pp interactions at $\sqrt{s} = 13$\TeV and decay exclusively via the $\bW$
channel have been presented. Events are selected requiring that one $\PW$
boson decays to a lepton and neutrino, and the other to a quark-antiquark
pair. The selection requires a muon or electron, significant
missing transverse momentum, and at least four jets. A kinematic fit assuming
$\TTbar$ or $\YYbar$ production is performed and for every event a candidate
$\T/\Y$ quark mass $\Mreco$ is reconstructed.
The analysis provides a high-resolution (7\%) mass scan of the \bW spectrum in the range from the top quark mass up to $\approx$2\TeV, in which
the signal from pair production of equal mass objects decaying to \bW would show up in a model-independent way as a narrow peak.
A binned maximum-likelihood fit to the $\Mreco$ distribution is made and
no significant deviations from the standard model expectations are found.
Upper limits are set on the $\TTbar$ and $\YYbar$ pair
production cross sections as a function of their mass.
By comparing these limits with the predicted theoretical cross section of
the pair production, the production of $\T$ and $\Y$ quarks is excluded
at 95\% confidence level for masses below 1295\GeV\,(1275\GeV expected).
More generally, the results set upper limits on the product of the
production cross section and branching fraction to $\bW$ for any new heavy
quark decaying to this channel.

\begin{acknowledgments}
We congratulate our colleagues in the CERN accelerator departments for the excellent performance of the LHC and thank the technical and administrative staffs at CERN and at other CMS institutes for their contributions to the success of the CMS effort. In addition, we gratefully acknowledge the computing centers and personnel of the Worldwide LHC Computing Grid for delivering so effectively the computing infrastructure essential to our analyses. Finally, we acknowledge the enduring support for the construction and operation of the LHC and the CMS detector provided by the following funding agencies: BMWFW and FWF (Austria); FNRS and FWO (Belgium); CNPq, CAPES, FAPERJ, and FAPESP (Brazil); MES (Bulgaria); CERN; CAS, MoST, and NSFC (China); COLCIENCIAS (Colombia); MSES and CSF (Croatia); RPF (Cyprus); SENESCYT (Ecuador); MoER, ERC IUT, and ERDF (Estonia); Academy of Finland, MEC, and HIP (Finland); CEA and CNRS/IN2P3 (France); BMBF, DFG, and HGF (Germany); GSRT (Greece); OTKA and NIH (Hungary); DAE and DST (India); IPM (Iran); SFI (Ireland); INFN (Italy); MSIP and NRF (Republic of Korea); LAS (Lithuania); MOE and UM (Malaysia); BUAP, CINVESTAV, CONACYT, LNS, SEP, and UASLP-FAI (Mexico); MBIE (New Zealand); PAEC (Pakistan); MSHE and NSC (Poland); FCT (Portugal); JINR (Dubna); MON, RosAtom, RAS, RFBR and RAEP (Russia); MESTD (Serbia); SEIDI, CPAN, PCTI and FEDER (Spain); Swiss Funding Agencies (Switzerland); MST (Taipei); ThEPCenter, IPST, STAR, and NSTDA (Thailand); TUBITAK and TAEK (Turkey); NASU and SFFR (Ukraine); STFC (United Kingdom); DOE and NSF (USA).

\hyphenation{Rachada-pisek} Individuals have received support from the Marie-Curie program and the European Research Council and Horizon 2020 Grant, contract No. 675440 (European Union); the Leventis Foundation; the A. P. Sloan Foundation; the Alexander von Humboldt Foundation; the Belgian Federal Science Policy Office; the Fonds pour la Formation \`a la Recherche dans l'Industrie et dans l'Agriculture (FRIA-Belgium); the Agentschap voor Innovatie door Wetenschap en Technologie (IWT-Belgium); the Ministry of Education, Youth and Sports (MEYS) of the Czech Republic; the Council of Science and Industrial Research, India; the HOMING PLUS program of the Foundation for Polish Science, cofinanced from European Union, Regional Development Fund, the Mobility Plus program of the Ministry of Science and Higher Education, the National Science Center (Poland), contracts Harmonia 2014/14/M/ST2/00428, Opus 2014/13/B/ST2/02543, 2014/15/B/ST2/03998, and 2015/19/B/ST2/02861, Sonata-bis 2012/07/E/ST2/01406; the National Priorities Research Program by Qatar National Research Fund; the Programa Severo Ochoa del Principado de Asturias; the Thalis and Aristeia programs cofinanced by EU-ESF and the Greek NSRF; the Rachadapisek Sompot Fund for Postdoctoral Fellowship, Chulalongkorn University and the Chulalongkorn Academic into Its 2nd Century Project Advancement Project (Thailand); the Welch Foundation, contract C-1845; and the Weston Havens Foundation (USA).
\end{acknowledgments}

\bibliography{auto_generated}
\cleardoublepage \appendix\section{The CMS Collaboration \label{app:collab}}\begin{sloppypar}\hyphenpenalty=5000\widowpenalty=500\clubpenalty=5000\textbf{Yerevan Physics Institute,  Yerevan,  Armenia}\\*[0pt]
A.M.~Sirunyan, A.~Tumasyan
\vskip\cmsinstskip
\textbf{Institut f\"{u}r Hochenergiephysik,  Wien,  Austria}\\*[0pt]
W.~Adam, F.~Ambrogi, E.~Asilar, T.~Bergauer, J.~Brandstetter, E.~Brondolin, M.~Dragicevic, J.~Er\"{o}, M.~Flechl, M.~Friedl, R.~Fr\"{u}hwirth\cmsAuthorMark{1}, V.M.~Ghete, J.~Grossmann, J.~Hrubec, M.~Jeitler\cmsAuthorMark{1}, A.~K\"{o}nig, N.~Krammer, I.~Kr\"{a}tschmer, D.~Liko, T.~Madlener, I.~Mikulec, E.~Pree, D.~Rabady, N.~Rad, H.~Rohringer, J.~Schieck\cmsAuthorMark{1}, R.~Sch\"{o}fbeck, M.~Spanring, D.~Spitzbart, W.~Waltenberger, J.~Wittmann, C.-E.~Wulz\cmsAuthorMark{1}, M.~Zarucki
% ...
\vskip\cmsinstskip
% ...
1:~~Also at Vienna University of Technology, Vienna, Austria\\
% ...

\end{sloppypar}
\end{document}